\documentclass[%
 reprint,
superscriptaddress,
showpacs,
 amsmath,amssymb,
 aps,
prl,
Ifloatfix,
citeautoscript,
]{revtex4-1}
\usepackage{CJK}
\usepackage[UKenglish]{babel}
\usepackage{graphicx}
\usepackage{epstopdf}
\usepackage{mathrsfs}
\usepackage{breakcites}
\usepackage{float}
\usepackage{tensor}
\usepackage{siunitx}
\usepackage{graphicx}
\usepackage{dcolumn}
\usepackage{bm}
\usepackage[normalem]{ulem}
\usepackage{color}

\providecommand{\diff}[1]{\ensuremath{\mathrm{d}{#1}}}
\providecommand{\rbracket}[1]{\ensuremath{\left( #1 \right)}}
\providecommand{\frbracket}[1]{\ensuremath{\!\left( #1 \right)}}
\providecommand{\sbracket}[1]{\ensuremath{\left[ #1 \right]}}

\providecommand{\cbracket}[1]{\ensuremath{\left\{ #1 \right\}}}

\usepackage{tensor}
\usepackage{braket}

\begin{document}
\begin{CJK*}{UTF8}{}

\preprint{APS/123-QED}

\title{Black Hole Field Theory with a Firewall in two spacetime dimensions}
\CJKfamily{bsmi}
\author{C. T. Marco Ho \CJKkern(何宗泰)}
\email{c.ho1@uq.edu.au}
\affiliation{Centre for Quantum Computation and Communication Technology, School of Mathematics and Physics, University of Queensland, Brisbane, Queensland 4072, Australia}

\author{Daiqin Su \CJKkern(粟待欽)}
\affiliation{Centre for Quantum Computation and Communication Technology, School of Mathematics and Physics, University of Queensland, Brisbane, Queensland 4072, Australia}

\author{Robert B. Mann}
\email{rbmann@uwaterloo.ca}
\affiliation{Centre for Quantum Computation and Communication Technology, School of Mathematics and Physics, University of Queensland, Brisbane, Queensland 4072, Australia}
\affiliation{Department of Physics and Astronomy, University of Waterloo, Waterloo, Ontario N2L 3G1, Canada}
\affiliation{Perimeter Institute, 31 Caroline Street North, Waterloo, Ontario N2L 2Y5, Canada}

\author{Timothy C. Ralph}%
 \email{ralph@physics.uq.edu.au}
\affiliation{Centre for Quantum Computation and Communication Technology, School of Mathematics and Physics, University of Queensland, Brisbane, Queensland 4072, Australia}

\date{\today}

\begin{abstract}
We propose that the vacuum state of a scalar field around a black hole is a modified Unruh vacuum. In (1+1) dimensions, we show that a free-faller close to such an horizon can be modelled as an inertial observer in a modified Minkowski vacuum. The modification allows for information-leaking correlations at high frequencies. Using a Gaussian detector centred at $k_0$, we find that the expectation value of the number operator for a detector crossing the horizon is proportional to $1/|k_0|$, implying that the free-faller will observe unbounded numbers of high energy photons, i.e. a firewall.

\end{abstract}
\maketitle
\end{CJK*}
The black hole information paradox, unresolved since its proposal by Hawking \cite{hawking_breakdown_1976}, stems from the thermality of Hawking radiation. Assuming that Hawking radiation continues until the black hole is completely evaporated, the process of black hole evaporation could change a pure state into a mixed state, which is forbidden under unitary evolution. Of course, given the lack of a consonant theory of gravity and quantum mechanics, almost all aspects of black hole evolution can be questioned \cite{hossenfelder_conservative_2010}. There are modified theories/suggestions where there is no black hole information paradox to resolve -- such as remnants \cite{aharonov_unitarity_1987,banks_black_1993} and information loss \cite{hawking_breakdown_1976} --
but these generate further difficulties and do not satisfactorily resolve the problem \cite{mann_black_2015}.
Black hole complementarity \cite{susskind_stretched_1993} was one attempt to resolve the information paradox. Whilst problems with complimentary had been raised earlier \cite{braunstein_quantum_2007}, the discord within complementarity was recently thrown into sharp focus by Almheiri et al. \cite{almheiri_black_2013} who found that the postulates of black hole complementarity: unitarity, the equivalence principle, and field-theory locality are mutually inconsistent. Almheiri et al. suggested that the most conservative solution was to forgo the equivalence principle such that a free falling observer will not observe a vacuum state and burns up in a `firewall' at the horizon. In a related and supporting development \cite{braunstein_better_2013}, Braunstein, Pirandola and \.{Z}yczkowski considered entanglement across the horizon which is `disentangled' near the end of life of a black hole, finding that this resulted in a firewall-like `energetic curtain'. In this paper we investigate the conservative solution to the firewall paradox of Almheiri et al., offering a mechanism for the firewall.  In contrast to a recent model \cite{louko_unruh-dewitt_2014,martin-martinez_$1+1mathrmd$_2015} that imposed breaking of correlations across a horizon (whose physical mechanism remains to be determined), we propose a field theoretical approach that entails choosing a different vacuum state. 

Which vacuum state though? The standard choice for the vacuum state around an eternal black hole is the Unruh vacuum. However, it is well known \cite{candelas_vacuum_1980,israel_black_2003,wald_quantum_1994} that in addition to the Unruh vacuum, there are also the Hartle-Hawking and Boulware vacua. Indeed, a generic result of quantum field theory is that there is no unique vacuum state. There are many complete sets of modes and each set is associated with a vacuum state. These vacuum states are not all equivalent, so there are a large number of vacua that can be defined. This proliferation of vacua means that for each physical situation, we require additional constraints to determine a vacuum state.

Here we suggest that an appropriate alternative vacuum choice would be a modified Unruh vacuum, one that transitions from an Unruh vacuum at low frequency to a Boulware vacuum \cite{boulware_quantum_1975} at high frequency. 
By exploiting the correspondence \cite{thorne_black_1986,fuentes-schuller_alice_2005} between the Rindler  and Schwarzschild spacetimes close to the horizon, we compute the response of a free falling detector across the horizon of a (1+1) dimensional Schwarzschild black hole. Given the choice of a modified Unruh vacuum we identify conditions for which the free-faller sees a firewall, whilst a stationary observer sees Hawking radiation--as proposed by Almheiri et al. \cite{almheiri_black_2013}.

\textit{Boulware and Unruh vacua, and Black Holes}---According to Israel \cite{israel_black_2003} and others \cite{wipf_quantum_1998,fabbri_semiclassical_2006}, the Boulware vacuum state is, {\itshape``...the zero-temperature ground state appropriate to the space in and around a static star.''}  Although black holes (excluding primordials) form from collapsing stars \cite{celotti_astrophysical_1999}, the derivation of Hawking radiation using an eternal black hole \cite{almheiri_black_2013,mann_black_2015,christensen_trace_1977,unruh_notes_1976} asserts that the vacuum state around the black hole is the Unruh vacuum. A particularly subtle point is illustrated by comparing the different derivations of Hawking radiation. The first, which we shall call the collapse model, was pioneered by Hawking \cite{hawking_particle_1975} and the second, which we shall call the eternal model, is exemplified by Unruh's paper \cite{unruh_notes_1976}. 

In the collapse model only one definition of vacuum is considered 
\cite{hawking_particle_1975,wald_particle_1975}, namely it is the state annihilated by  asymptotic modes at past null infinity $\mathscr{I}^-$ that are positive frequency with respect to Schwarzschild time. This definition is also used in defining the Boulware and Unruh vacua in the eternal model. However in the collapse model this specification completely defines the vacuum, whereas in the eternal case additional initial conditions must be specified.  We therefore posit  that the vacuum state of the field may not be simply Boulware or Unruh, but something in between. We shall call this a modified Unruh vacuum.

In the eternal black hole case \cite{unruh_notes_1976},
\begin{equation}
\Ket{0_U} = Z\prod_{\omega,l,m} \text{exp}\cbracket{e^{-4\pi M \omega} \tensor*[^-_+]{{b}}{_\omega_l_m^\dagger} \tensor*[^-_-]{{b}}{_\omega_l_m^\dagger}}\Ket{0_B}, \label{eq:unruhvacuum}
\end{equation}
where $\Ket{0_U}$ is the Unruh vacuum while $\Ket{0_B}$ is the Boulware vacuum and Z is a normalisation constant. $M$ is the mass of the Schwarzschild black hole, $\omega$ is the frequency measured with respect to Schwarzschild time, $l$ an $m$ are the degree and order in the spherical harmonic function $Y_l^m$. $\tensor*[^-_+]{{b}}{_\omega_l_m^\dagger} $ $\rbracket{\tensor*[^-_-]{{b}}{_\omega_l_m^\dagger}}$ creates a positive frequency mode with respect to Schwarzschild time $t$ in the right (left) exterior region of the maximally extended Schwarzschild solution that is asymptotically outgoing at spatial infinity. The pre-subscript indicates the exterior region while the negative pre-superscripts indicate the modes are outgoing in the exterior regions. The right exterior region is identified with our universe and the left exterior outgoing mode is interpreted to be the `particle' inside the horizon \cite{unruh_notes_1976} or in the terminology of Almheiri et al., the `ingoing' mode \footnote{Why the left exterior \emph{outgoing} mode is considered to be the \emph{ingoing} mode is due to the analysis using the eternal black hole. In the collapse scenario there is no left exterior region and the left exterior region outgoing mode is replaced by an ingoing (into the black hole) mode that Wald has called the `particles which go down the black hole' \cite{unruh_notes_1976,wald_particle_1975}.}. 

The modification to the Unruh vacuum in Eq.~\eqref{eq:unruhvacuum} that we propose is
\begin{equation}
\Ket{\tilde{\Psi}} = Z\prod_{\omega<\tilde{\epsilon},l,m} \text{exp}\cbracket{e^{-4\pi M \omega} \tensor*[^-_+]{{b}}{_\omega_l_m^\dagger} \tensor*[^-_-]{{b}}{_\omega_l_m^\dagger}}\Ket{0_B}, \label{eq:modifiedboulware}
\end{equation}
where $\tilde{\epsilon}$ is a cutoff frequency. This modification ensures Hawking radiation is retained for  frequencies up to $\tilde{\epsilon}$, with higher frequencies available to carry out information via a modified theory. In what follows we shall explore the implications of this in Rindler space, modifying Eq.~\eqref{eq:modifiedboulware} accordingly.

\textit{Correspondence between Schwarzschild and Rindler metrics}---We propose that the vacuum state after gravitational collapse is a modified Unruh vacuum state, and are particularly interested in the number of particles that a freely falling observer would detect near the horizon. To compute this we shall exploit  the well known correspondence between Schwarzschild and Rindler metrics \cite{thorne_black_1986}, which has been useful in relativistic quantum information \cite{fuentes-schuller_alice_2005}. 
Specifically, near the event horizon
\begin{figure}
\centering
\includegraphics[width=\columnwidth]{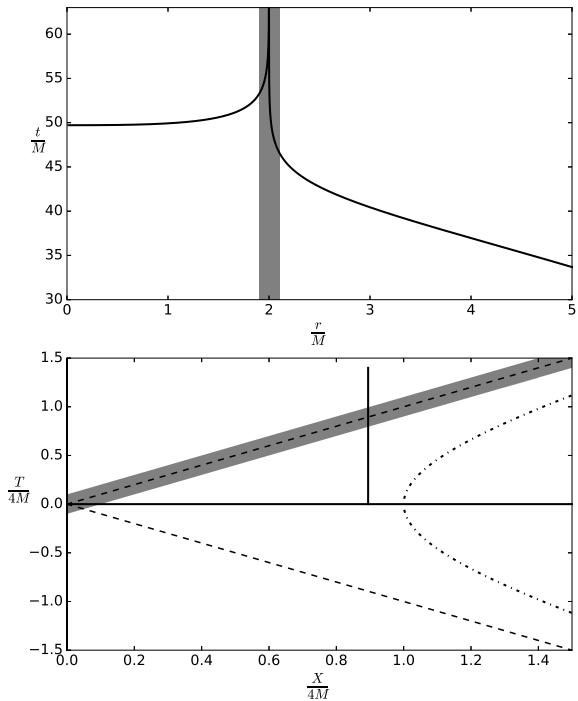}
\caption{The top figure plots the geodesic initially stationary at $r = 10 M$; note that the divergence is an artifact of Schwarzschild coordinates and is absent in the free-faller's coordinates. The bottom figure shows the corresponding geodesic in flat space. The dot-dashed line is the $\xi = 0$ trajectory. The shaded region on both plots indicates the area where the flat space equivalence holds.} \label{fig:compare}
\end{figure}
\begin{equation}
{\mathrm{d}s}^2 = \rbracket{1-\frac{2M}{r}} {\mathrm{d}t}^2 - \frac{1}{1-\frac{2M}{r}} {\mathrm{d}r}^2 \approx e^{2a\xi} \rbracket{\diff{\eta}^2- \diff{\xi}^2},
\end{equation}
where $t$ is Schwarzschild time, $r$ is the Schwarzschild radial coordinate, $\eta$ is Rindler time and $\xi$ is the Rindler space coordinate. The approximate relation follows from setting $t=\eta$, $r=2M \rbracket{1+e^{2a\xi}}$ and expanding for $\xi\ll 0$, where $a = \frac{1}{4M}$.
The relationship between Rindler and Minkowski coordinates $\rbracket{T,X}$ is
\begin{equation}
T = \frac{e^{a\xi}}{a}\sinh{(a\eta)}, \quad  X = \frac{e^{a\xi}}{a}\cosh{(a\eta)} \quad X>|T|
\label{eq:rindlerdef1}
\end{equation}
\begin{equation}
T = -\frac{e^{a\xi}}{a}\sinh{(a\eta)}, \quad  X = -\frac{e^{a\xi}}{a}\cosh{(a\eta)} \quad X<-|T|.
\label{eq:rindlerdef2}
\end{equation}

Henceforth we can always associate $\xi$ with a radius $r$ from a black hole. It must be noted that the correspondence is only approximate and holds with greater accuracy as $\xi$ gets more negative. We see in Fig.~\ref{fig:compare} that as $\xi$ gets more negative, we get closer to $r=2M$, the event horizon which corresponds to $T = \pm X$.

\textit{Motion of the falling observer}---For a freely falling (and initially stationary at $r_0$) observer,
\begin{equation}
\frac{\diff{x^\mu}}{\diff{\tau}} =\rbracket{\frac{\sqrt{1-\frac{2M}{r_0}}}{1-\frac{2M}{r}},-\sqrt{\frac{2M}{r} - \frac{2M}{r_0}},0,0}.
\label{eq:schwarzschildfaller}
\end{equation}

We shall now identify a Minkowski inertial trajectory with a the freely falling observer's trajectory in Schwarzschild. Consider a trajectory such that $X_0 = \text{const}$. From Eq.~\eqref{eq:rindlerdef1} we have $X_0^2 - T^2 = \frac{e^{2a\xi}}{a^2}$. We rearrange this, make the approximation that $\frac{e^{2a\xi}}{X_0^2 a^2}\ll 1$ and use $r=2M \rbracket{1+e^{2a\xi}}$ to get,
\begin{equation}
T \approx  X_0\rbracket{1- \frac{1}{2 X_0^2 a^2} \rbracket{\frac{r}{2M}-1}}.
\label{eq:inertialapprox}
\end{equation}
We then integrate $\frac{\diff{r}}{\diff{\tau}} = -\sqrt{\frac{2M}{r} - \frac{2M}{r_0}}$ in Eq.~\eqref{eq:schwarzschildfaller} and then series expand around $r = 2M$ to get,
\begin{equation}
\tau \approx \text{const} - \frac{r-2M}{\sqrt{1-\frac{2M}{r_0}}}.
\label{eq:integrationapprox}
\end{equation}
If we now identify Eq.~\eqref{eq:inertialapprox} with Eq.~\eqref{eq:integrationapprox}, we may compare the coefficient of $r$ to obtain,
\begin{equation}
X_0 = 4M \sqrt{\frac{r_0-2M}{r_0}}.
\label{eq:inertialtrajectory}
\end{equation}

\textit{Number expectation for an observer coupling to Minkowski modes in the modified Minkowski vacuum}---To calculate the response of a free falling detector we need to identify an appropriate vacuum state. The outgoing modes in the right exterior region in the Unruh vacuum of the standard approach map to the right moving modes of the Minkowski vacuum in the equivalent Rindler metric. Thus an inertial observer (free-faller) near the horizon sees no particles coming from the horizon, in accord with the equivalence principle. We note that such a observer does see non-zero particles from the sky (see discussion after Eq. (3.19) in \cite{unruh_notes_1976}) and an Unruh-Dewitt detector has been shown \cite{juarez-aubry_onset_2014} to respond non-trivially  throughout its free fall in (1+1) Schwarzschild spacetime. On the other hand, a stationary observer in the Schwarzschild metric, with the Boulware vacuum, sees no particles. The equivalent vacuum for the Rindler metric is the Rindler vacuum for which accelerated observers see no particles. 

We now consider what an observer measuring in Minkowski modes will detect in a modified vacuum, focussing on a massless scalar field. The Minkowski vacuum can formally be written in terms of the Rindler vacuum \cite{takagi_vacuum_1986},
\begin{align}
\Ket{0_M} \propto \text{exp}\cbracket{\frac{1}{2}{\int_0^\infty\diff{\Omega}\sum_{\sigma} {e^{-\pi\Omega}} {b_{\Omega}^{\rbracket{\sigma}}}^\dagger {b_{\Omega}^{\rbracket{-\sigma}}}^\dagger}}\Ket{0_R}, \label{eq:RtoM}
\end{align}
 corresponding to Eq.~\eqref{eq:unruhvacuum}, with $\Omega$   the Rindler frequency and $\sigma = 1$ $\rbracket{\sigma = -1}$ signifying the right (left) Rindler wedge, where  $b_{\Omega}^{\rbracket{\sigma}}$ are Rindler modes. 

To implement our proposed modified Unruh vacuum Eq.~\eqref{eq:modifiedboulware}, we suppose that for Rindler frequencies below $\epsilon$ the state appears to be Minkowski vacuum and therefore define \begin{align}
\Ket{\Psi} \propto \text{exp}\cbracket{\frac{1}{2}{\int_0^\epsilon \diff{\Omega} \sum_{\sigma} e^{-\pi\Omega} {b_{\Omega}^{\rbracket{\sigma}}}^\dagger {b_{\Omega}^{\rbracket{-\sigma}}}^\dagger}}\Ket{0_R}
\label{eq:modifiedRindlervacuum}
\end{align}
which is a modified Minkowski vacuum. Note that we have suppressed an index distinguishing left and right movers in Eqs.~\eqref{eq:RtoM} and \eqref{eq:modifiedRindlervacuum}.

One might expect based on the definition of the Unruh vacuum in Eq. \eqref{eq:unruhvacuum} that the left movers in Eq. \eqref{eq:RtoM} and \eqref{eq:modifiedRindlervacuum} would be in the Rindler vacuum since they correspond to the ingoing modes outside the black hole which are in the Boulware vacuum. This is not the case in (3+1) dimensions.  As was discussed in \cite{crispino_comment_2012,singleton_singleton_2012} and shown in  \cite{candelas_vacuum_1980}, in the (3+1) dimensional case an Unruh-Dewitt detector hovering above the horizon in the Unruh vacuum has a response function identical to that of a uniformly accelerated observer in the Minkowski vacuum. This is in contrast to the (1+1) dimensional case where the correspondence breaks down due to the decoupling of left and right movers. 
Physically in the (3+1) case the response is due to the sum over all angular momentum modes. If we are near the horizon and are not looking directly radially away, most of the modes we see are outgoing modes that have been bent back towards the black hole. Indeed, it has been shown \cite{candelas_vacuum_1980} that ingoing modes can be neglected and an Unruh-Dewitt detector near the horizon sees a response only due to the outgoing modes. Because of the bending of the outgoing modes, the detector hovering at the horizon sees isotropic radiation as in the Unruh effect. This suggests that for a (1+1) theory to model what a detector in an Unruh vacuum near the horizon of a (3+1) black hole sees, the state of the field should be in the Minkowski vacuum  \eqref{eq:RtoM} for both left and right movers. Likewise, for the modified Unruh vacuum \eqref{eq:modifiedboulware} the flat space analogue is the modified Minkowski vacuum state  \eqref{eq:modifiedRindlervacuum} for both left and right movers. In this way, in the limit of $\epsilon \rightarrow \infty$ we regain the standard correspondence.

We see that for Rindler operators ${b_{\Omega'}^{\rbracket{\sigma}}}$ with $\Omega'<\epsilon$, Eq.~\eqref{eq:modifiedRindlervacuum}  behaves like a Minkowski vacuum while for $\Omega'>\epsilon$ it behaves like a Rindler vacuum. This is the equivalent of our modified Unruh vacuum, valid in (1+1) when we are close to the horizon.

Note that $\Omega$ is the Rindler frequency scaled with respect to $a$, therefore $\Omega = \frac{\omega}{a}$ where $\omega$ is the Rindler frequency measured with respect to the Rindler time $\eta$ in Eq.~\eqref{eq:rindlerdef1}.

If we now calculate the number expectation value for a Minkowski mode $a_{k}$, with $\Ket{\Psi}$ instead of $\Ket{0_R}$ we find that
\begin{align}
&\Braket{\Psi|{a_{{k}}}^\dagger{a_{{k}}}|\Psi} \notag\\
 &= \frac{1}{2\pi\omega_{k}} \int_{\epsilon}^\infty \diff{\Omega} \frac{2}{e^{2\pi\Omega}-1}\notag \\
&= \frac{1}{\pi\omega_{k} }  \sbracket{\epsilon - \frac{1}{2\pi}\text{ln}\rbracket{e^{2\pi\epsilon}-1}}, \label{eq:singlefreq}
\end{align}
where $\omega_{k} = |k|$.

\begin{figure}
	\centering
	\includegraphics[width=\columnwidth]{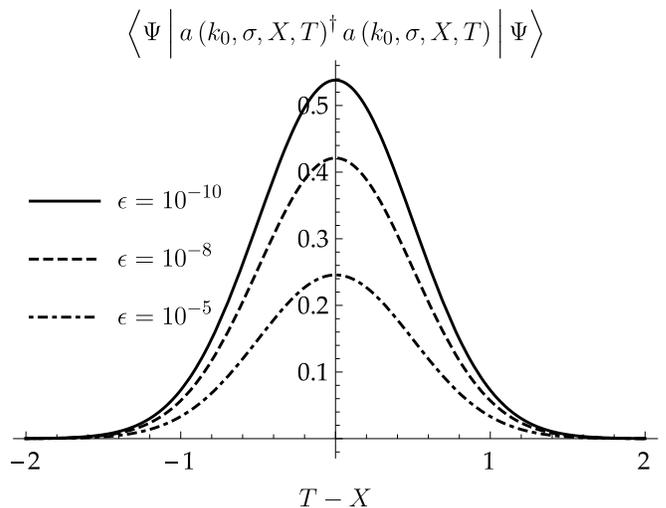}
	\caption{\label{fig:numberexpect} Numerical integration of Eq.~\eqref{eq:numberexpect} with $\sigma = 1$ and $k_0 = 10$ plotted with respect to $T-X$. We see that as we approach the horizon at $T-X = 0$, we encounter a finite response due to the firewall. Notice our approximations above Eq. \eqref{eq:inertialapprox} mean that this figure is valid only for $T-X \ll 2M$.}
\end{figure}

\textit{Position dependent response to the firewall}--- Consider the superposition of Minkowski modes $a_k$
\begin{equation}
a\frbracket{k_0,\sigma,X,T} = \int_{-\infty}^\infty \diff{k} ~f\frbracket{k,k_0,\sigma,X,T} a_{k} \label{eq:adef}
\end{equation}
with  $ f\frbracket{k, k_0,\sigma,X,T}=\rbracket{\frac{1}{2\pi \sigma^2}}^{\frac{1}{4}}e^{-\frac{\rbracket{k-k_0}^2}{4\sigma^2}} e^{-i \rbracket{\omega_k T -k X}}$. This wave-packet models a detector of finite size measuring with respect to a Gaussian wave-packet of Minkowski modes centred at $k_0$ with a width of $\sigma$. Assuming $|k_0|\gg \sigma$ we  find 
\begin{align}
&\Braket{\Psi| a\frbracket{k_0,\sigma,X,T}^\dagger a\frbracket{k_0,\sigma,X,T}|\Psi}\notag \\
&=\int_\epsilon^\infty \diff{\Omega} \frac{2\sigma}{|k_0| \sqrt{2\pi}} \frac{1}{e^{2\pi \Omega}-1}\notag \\
&\qquad \times \sbracket{e^{-2\frac{\sigma^2 \sbracket{\Omega-\rbracket{|k_0|T- k_0X}}^2}{{k_0}^2}} +e^{-2\frac{\sigma^2 \sbracket{\Omega+\rbracket{|k_0|T-k_0X}}^2}{{k_0}^2}}} \label{eq:numberexpect}
\end{align}
for the  approximate position-dependent
expectation value of the number operator in the modified Rindler vacuum. 
We plot this quantity in Fig.~\ref{fig:numberexpect}  for parameters $k_0 = 10$, $\sigma = 1$ and for three different values of $\epsilon $.  This result is a generalisation of \cite{louko_transition_2008} where they calculate the response of an inertial Unruh-Dewitt detector in the Rindler vacuum. With our detector model, this corresponds to setting $\epsilon = 0$ and $X = \text{constant}$.

\begin{figure}
\centering
\includegraphics[width=\columnwidth]{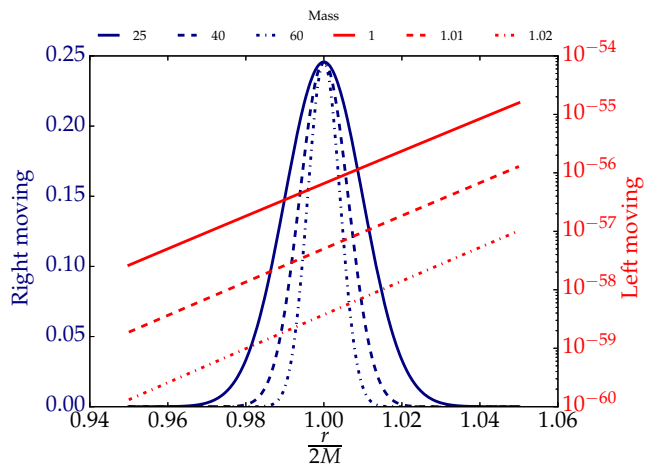}
\caption{\label{fig:leftright} Numerical integration of Eq.~\eqref{eq:numberexpect} with $\sigma = 1$ and  right movers ($k_0 = 10$) in blue and left movers ($k_0=-10)$ in red plotted with respect to $\frac{r}{2M}$ using Eq. \eqref{eq:inertialapprox} and \eqref{eq:inertialtrajectory} with $\epsilon=10^{-5}$ and for several values of black hole mass $M$. Notice the large differences in the particle flux coming from the horizon (right movers) versus that coming from the sky (left-movers) for the free-faller.}
\end{figure}

If we now consider a free falling observer dropped from $r_0$, their inertial trajectory is given by Eq.~\eqref{eq:inertialtrajectory} with $X=X_0$. To satisfy the semi-classical assumption \cite{almheiri_black_2013}, $M$ must be large, so we assume that $M\gg0$. The response of such an observer, coupling to Minkowski modes can be seen in Fig.~\ref{fig:numberexpect}. If the detector measures right-moving modes ($k_0>0$), then as they approach the horizon at $T = X$ they encounter a finite number of particles. $X$ is very large for $M\gg0$ so $T+X$ is very large, therefore, detectors measuring left-moving modes ($k_0<0$) see very few particles as compared to right movers. Furthermore the number of particles they see from behind (i.e. left-movers) is much smaller for larger mass black holes (See Fig. \ref{fig:leftright}). This means that the freely falling observer runs into a `firewall' at the horizon with negligible particles from behind as is expected.
Interestingly, this firewall is not thermal. This can be most easily seen in Eq.~\eqref{eq:singlefreq} and Eq.~\eqref{eq:numberexpect}. The dependence on frequency at the horizon is $\frac{1}{|k_0|}$.

The particle number is larger for smaller $\epsilon$ which can be most easily seen in Eq.~\eqref{eq:singlefreq}. However, even for large $\epsilon$, the inverse dependence on $k_0$ means that the number of particles above any frequency is unbounded at the horizon in the absence of an UV cutoff. For example, suppose we consider a hydrogen atom in the ground state. There are an unbounded number of photons at and above the ionization energy $\SI{13.6}{\electronvolt}$ and so any hydrogen atom falling into the black hole will be ionized. This will be true for all atoms and molecules.

\textit{Cutoff}---The correlations between the two modes, ${b_{\Omega}^{\rbracket{\sigma}}}$ and ${b_{\Omega}^{\rbracket{-\sigma}}}$ in Eq.~\eqref{eq:RtoM} is both the cause of Hawking radiation and the origin of the problem raised by the firewall argument. In Eq.~\eqref{eq:modifiedRindlervacuum} we have excised the correlations between the two modes for $\Omega > \epsilon$. This allows the higher frequency modes to be correlated (via some modification of the standard theory) with other modes---as is required for information leakage---but has the unfortunate effect of eliminating Hawking radiation at these frequencies. 

The problem is to determine $\epsilon$. We note that via Wien's displacement law, the frequency spectral radiance of black body radiation peaks at $\omega_\text{peak} \approx 2\pi \times 2.82144 T$. The cutoff frequency must therefore be $\omega_\text{cutoff} = \epsilon a \gg \omega_\text{peak}$. Using the Hawking temperature, this imposes the lower bound $\epsilon \gg 2.82144$, ensuring that most of the energy is given out via Hawking radiation for all but the highest frequencies. The specification of the cutoff at $\Omega_\text{cutoff} = \epsilon$ is independent of $a$.

\textit{Squeezing}---An interesting property of the modified Minkowski vacuum is that Minkowski modes are squeezed. We define the $X$ quadrature as $X = a + a^\dagger$ and the $P$ quadrature as $P = -i\rbracket{a-a^\dagger}$, where $a$ is given by Eq.~\eqref{eq:adef} with $T=X=0$. $\rbracket{\Delta X}^2$ is defined as $\rbracket{\Delta X}^2 = {\Braket{\Psi|X^2|\Psi} - (\Braket{\Psi|X|\Psi}})^2$. $\Delta P$ is defined similarly.  To simplify calculations, we can apply the approximation $|k| \gg 0$ and $|k| \gg \sigma$. With these approximations we find the variances
\begin{align}
&\rbracket{\Delta X}^2 = 1 +\frac{2^{\frac{5}{2}}\sigma}{\pi^\frac{1}{2} |k_0|} \rbracket{\epsilon - \frac{\ln\rbracket{e^{\pi \epsilon}+1}}{\pi}}, \label{eq:sigmax}\\
&\rbracket{\Delta P}^2 = 1 + \frac{2^{\frac{5}{2}}\sigma}{\pi^\frac{1}{2} |k_0|} \rbracket{\epsilon - \frac{\ln\rbracket{e^{\pi \epsilon} -1}}{\pi}},
\end{align}
yielding
\begin{align}
&\lim_{\epsilon \rightarrow 0 } \rbracket{\Delta X}^2 = 1 - \frac{2^{\frac{5}{2}} \sigma\ln 2}{\pi^\frac{3}{2} |k_0|}, \label{eq:sigmaxlimit}\\
&\lim_{\epsilon \rightarrow 0 } \rbracket{\Delta P}^2\rightarrow \frac{2^{\frac{5}{2}} \sigma}{\pi^\frac{3}{2} |k_0|} \ln\rbracket{\epsilon} \rightarrow \infty
\end{align}
for $\epsilon \rightarrow 0$, in the limit of the Rindler vacuum.
We see in Eq.~\eqref{eq:sigmax} and Eq.~\eqref{eq:sigmaxlimit}, $\rbracket{\Delta X}^2$ is always less than 1, indicating single-mode squeezing (though not purely single-mode). We also see that squeezing disappears for $|k_0| \rightarrow \infty$ or for $\epsilon \rightarrow \infty$.

The origin of the squeezing comes from the Rindler aspects of the vacuum state. While the Minkowski vacuum is not a squeezed state of Minkowski modes, the Rindler vacuum, Unruh modes $\cbracket{d_{\Omega}^{\rbracket{\sigma}}}$ can be shown to be two-mode squeezed,
\begin{equation}
\Ket{0_R} \propto \text{exp}\cbracket{-\frac{1}{2}\int\diff{\Omega} \sum_\sigma e^{-\pi\Omega} {d_{\Omega}^{\rbracket{\sigma}}}^\dagger {d_{\Omega}^{\rbracket{-\sigma}}}^\dagger}\Ket{0_M}. \label{eq:Unruhsqueezed}
\end{equation}
Unruh modes can be written in terms of Minkowski modes
\begin{equation}
d_{\Omega}^{\rbracket{\sigma}} = \int_{-\infty}^\infty \diff{k} ~ {p_\Omega^{\rbracket{\sigma}}\rbracket{k}} a_{k}.
\label{eq:UnruhMinkowski}
\end{equation}
Upon insertion of Eq.~\eqref{eq:UnruhMinkowski} into \eqref{eq:Unruhsqueezed}, we see that the Rindler vacuum is multi-mode squeezed in terms of Minkowski modes. Part of the multi-mode squeezing contains single-mode squeezing and it is this that contributes to the squeezing we see.

\textit{Correlations and leaking information}---Central to the firewall argument of Almheiri et al. \cite{almheiri_black_2013} is the entanglement---in the standard picture---between outgoing and ingoing modes of radiation due to the free falling observer seeing a vacuum. This is the entanglement between $\tensor*[^-_+]{{b}}{_\omega_l_m^\dagger}$ and $ \tensor*[^-_-]{{b}}{_\omega_l_m^\dagger}$ in the Unruh vacuum. The entanglement precludes the leaking of information out of the black hole, which requires that the outgoing early and late radiation form a pure state. In our proposed modified Unruh vacuum, a free faller does not see a vacuum; entanglement between any two high frequency modes is absent, thus allowing the possibility that these modes can be correlated in a more complicated theory such that information may be carried out.

\textit{Conclusion}---We have proposed a modified Unruh vacuum state to replace the Unruh vacuum state around a black hole. Using the correspondence between Schwarzschild and Rindler spacetimes close to the horizon, we constructed in (1+1) dimensions a modified Minkowski vacuum to model our modified Unruh state. Minkowski modes exhibit single-mode squeezing in this modified Minkowski vacuum state.

 We found that this vacuum state led to a firewall as predicted by Almheiri et al. \cite{almheiri_black_2013}. The firewall is strikingly non-thermal: the particle number for a Gaussian detector is inversely proportional to the frequency. Absent a UV cutoff, any atom or molecule will immediately be ionized by the firewall due to the unbounded numbers of high energy photons. Thus any entanglement carried by atoms will be destroyed. This is in contrast to the situation that imposes a breaking of correlations across the horizon 
\cite{louko_unruh-dewitt_2014,martin-martinez_$1+1mathrmd$_2015}. The physical mechanism  responsible for this remains to be found and it is not clear if it can satisfy unitarity requirements.

Our (1+1) dimensional calculation with the modified Minkowski  vacuum state \eqref{eq:modifiedRindlervacuum} is expected to be a very good approximation to the (3+1) dimensional case when a free falling observer detects very localised modes pointing in the radial direction. This provides us with confidence in our flat space calculation and the possibility that the standard theory with a modified vacuum choice could be enough to maintain Hawking radiation (with small deviations) and do the job of carrying out information. 
To better understand the implications of our proposal in more realistic settings will require a calculation of the behaviour of a  detector in our modified Unruh vacuum \eqref{eq:modifiedboulware} around an actual black hole. This is a project for future work.

 \medskip
 
 C.T.M.H. would like to thank Ben Duffus for discussions of value. This work was supported in part by the Natural Sciences and Engineering Research Council of Canada. This work was also partially supported by the Australian Research Council Centre of Excellence for Quantum Computation and Communication Technology (Project No. CE110001027).

\sloppy
\raggedright
\providecommand{\noopsort}[1]{}\providecommand{\singleletter}[1]{#1}%
%

\end{document}